\documentclass[runningheads]{llncs}
\usepackage{graphicx}
\usepackage[T1]{fontenc}
\usepackage{verbatimbox,lipsum}
\usepackage{subcaption}
\usepackage{listings}
\usepackage{amssymb}
\usepackage{pifont}
\newcommand{\cmark}{\ding{51}}%
\newcommand{\xmark}{\ding{55}}%
\usepackage{float}

\newfloat{lstfloat}{htbp}{lop}
\floatname{lstfloat}{Listing}
\begin{document}
\title{Enriching Scholarly Knowledge with Context}
%
%
\author{Muhammad Haris\inst{1}\orcidID{0000-0002-5071-1658} \and
Markus Stocker\inst{2}\orcidID{0000-0001-5492-3212} \and
Sören Auer\inst{2,1}\orcidID{0000-0002-0698-2864}}
\authorrunning{Haris et al.}
%
\institute{L3S Research Center, Leibniz University Hannover 30167, Hannover, Germany \\
\email{haris@l3s.de}\\
\and
TIB---Leibniz Information Centre for Science and Technology, Germany\\
\email{\{markus.stocker,auer\}@tib.eu}}
\maketitle              
\begin{abstract}
Leveraging a GraphQL-based federated query service that integrates multiple scholarly communication infrastructures (specifically, DataCite, ORCID, ROR, OpenAIRE, Semantic Scholar, Wikidata and Altmetric), we develop a novel web widget based approach for the presentation of scholarly knowledge with rich contextual information. We implement the proposed approach in the Open Research Knowledge Graph (ORKG) and showcase it on three kinds of widgets. First, we devise a widget for the ORKG paper view that presents contextual information about related datasets, software, project information, topics, and metrics. Second, we extend the ORKG contributor profile view with contextual information including authored articles, developed software, linked projects, and research interests. Third, we advance ORKG comparison faceted search by introducing contextual facets (e.g. citations). As a result, the devised approach enables presenting ORKG scholarly knowledge flexibly enriched with contextual information sourced in a federated manner from numerous technologically heterogeneous scholarly communication infrastructures.

\end{abstract}
\section{Introduction}
Massive (meta)data about digital and physical scholarly artefacts including articles, datasets, software, instruments, and samples are made available through various scholarly communication infrastructures~\cite{xia2017big,khan2017survey,Stocker_instruments}. Individually, current infrastructures focus on finding a certain kind of artefact. Lacking the ability to present information about related artefacts, they are unable to meet complex user information needs~\cite{safder2018ai}. For instance, if a researcher searches for scholarly articles she may want information about related datasets, software, projects and organizations. Obtaining such diverse information with a single request is not obvious because the information resides with distributed and technologically heterogeneous infrastructures. Separate search on infrastructures is, however, time consuming and laborious~\cite{6666765,Schwarte2011FedXOT}. Therefore, federated search is necessary for efficient and comprehensive content exploration.

For this purpose, we developed a GraphQL-based federated system~\cite{federated_access} that integrates multiple scholarly communication infrastructures, namely, the Open Research Knowledge Graph (ORKG)\footnote{\url{https://www.orkg.org/orkg/}}~\cite{orkg}, DataCite\footnote{\url{https://datacite.org/}}, and GeoNames\footnote{\url{https://www.geonames.org/}}. It supports executing queries in a federated manner and enables the integrated retrieval of scholarly information. The main purpose of the federated system is to enable cross-walking scholarly knowledge and contextual information as well as filtering at (meta)data granularity. However, the federated system currently has some limitations: 1) The scope of contextual information is limited to three scholarly infrastructures; and 2) the system requires queries to be written in GraphQL, which is untenable in practice.

As the main contribution of the work presented here, we devise a web widget based approach that retrieves rich contextual information for scholarly knowledge from distributed scholarly communication infrastructures and presents scholarly knowledge with rich context in an integrated manner. We demonstrate the integration of these widgets in ORKG to enrich its various views thus enabling rapid, comprehensive exploration of scholarly content. The proposed approach involves the following two main aspects:

\begin{enumerate}
    \item Extend the GraphQL-based federated system\footnote{\url{https://www.orkg.org/orkg/graphql-federated}} to include the DataCite PID Graph and REST APIs of OpenAIRE\footnote{\url{https://graph.openaire.eu/develop/api.html}}, Semantic Scholar\footnote{\url{https://www.semanticscholar.org/product/api}}, Wikidata\footnote{\url{https://www.wikidata.org/w/api.php}} and Altmetric\footnote{\url{https://api.altmetric.com/}} and enable retrieving comprehensive contextual information for ORKG scholarly knowledge in a federated and integrated manner.
    \item Building on the extended federated system, develop different web widgets to enrich scholarly knowledge viewed in ORKG with rich contextual information.
\end{enumerate}

We address the following research question: How can we flexibly enrich the presentation of scholarly knowledge in web based user interfaces with comprehensive contextual information published by numerous heterogeneous scholarly communication infrastructures?

\section{Related Work}
Scholia~\cite{Nielsen2017} is a dynamic user interface that operates on Wikidata's SPARQL endpoint and supports users in searching for articles, researcher profiles, organizations and publishers. Similarly, BioCarian~\cite{zaki2017} is a SPARQL endpoint powered user-friendly interface enabling exploring biological databases. The interface is enriched with facets that enable better query construction, thus making it easier for users to filter data. OSCAR ~\cite{Heibi2019} is a platform for searching RDF triples using a SPARQL endpoint while hiding the complexity of SPARQL, thus making the search operations easier for users who are not aware of web technologies. Similarly, Elda\footnote{\url{ http://www.epimorphics.com/web/tools/elda.html}} was proposed to access data served via a Linked Data API\footnote{\url{https://code.google.com/p/linkeddata-api}}. Elda is a Java implementation of the Linked Data API that allows customization of API requests for accessing RDF datasets. 

Following the Scholix~\cite{Scholix} framework, ScholeXplorer\footnote{\url{https://scholexplorer.openaire.eu/}} aggregates metadata harvested from different data sources (in particular, DataCite, Crossref, OpenAIRE) and creates a graph of scholarly entities. As such, the framework supports users in discovering research articles and related datasets. 

Kurteva and Ribaupierre~\cite{Kurteva2021} present a user interface that allows casual users to find specific types of data in the DBpedia knowledge base. The interface also provides a graphical visualization of retrieved results. Morton et al.~\cite{Morton2019} present a framework for querying biomedical knowledge graphs, ranking, and conveniently exploring the queried results. Several other systems for research data discovery exist including BioGraph~\cite{liekens2011biograph}, Het.io ~\cite{himmelstein2017systematic}, Wikidata\footnote{\url{https://www.wikidata.org/wiki/Wikidata:Main\_Page}}, Open Knowledge Maps\footnote{\url{https://openknowledgemaps.org/about}}, Unpaywall\footnote{\url{https://unpaywall.org/}}, Zenodo\footnote{\url{https://zenodo.org/}}, Figshare\footnote{\url{https://figshare.com/}}, re3data\footnote{\url{https://www.re3data.org/}}.

FedX~\cite{Schwarte2011FedXOT} was proposed to execute SPARQL queries on virtually integrated heterogeneous data sources. The practicability of the proposed framework was demonstrated by executing some real-world queries on the Linked Open Data Cloud. BioFed~\cite{Hasnain2017} is another federated query processing system that supports executing queries on a variety of SPARQL endpoints to retrieve life sciences data. The system integrates 130 SPARQL endpoints and supports retrieving the provenance information along with the data. The efficiency of the system was demonstrated by executing 10 complex and 10 simple queries, and the results were compared with FedX in terms of optimization. Another SPARQL-based federated system was proposed~\cite{federatedengine}, whose main purpose was to retrieve Open Educational Resources (OERs) published on disparate web platforms. Federated systems also support searching for personalized information, such as retrieving information about user profiles from diverse sources~\cite{arya}.

The structured comparison of different scholarly communication infrastructures can be found in Haris et al.~\cite{haris}. As the amount of data on these infrastructures is increasing rapidly, it is of utmost importance to enrich scholarly artefacts with their contextual information. The infrastructures reviewed here individually provide information about a particular kind of scholarly artefact, but rarely present the artefacts with rich contextual information. For ORKG scholarly knowledge as the core artefact, we propose an approach that queries a range of scholarly communication infrastructures to retrieve and present rich contextual information.

\begin{figure*}[t!]
  \includegraphics[width=\textwidth]{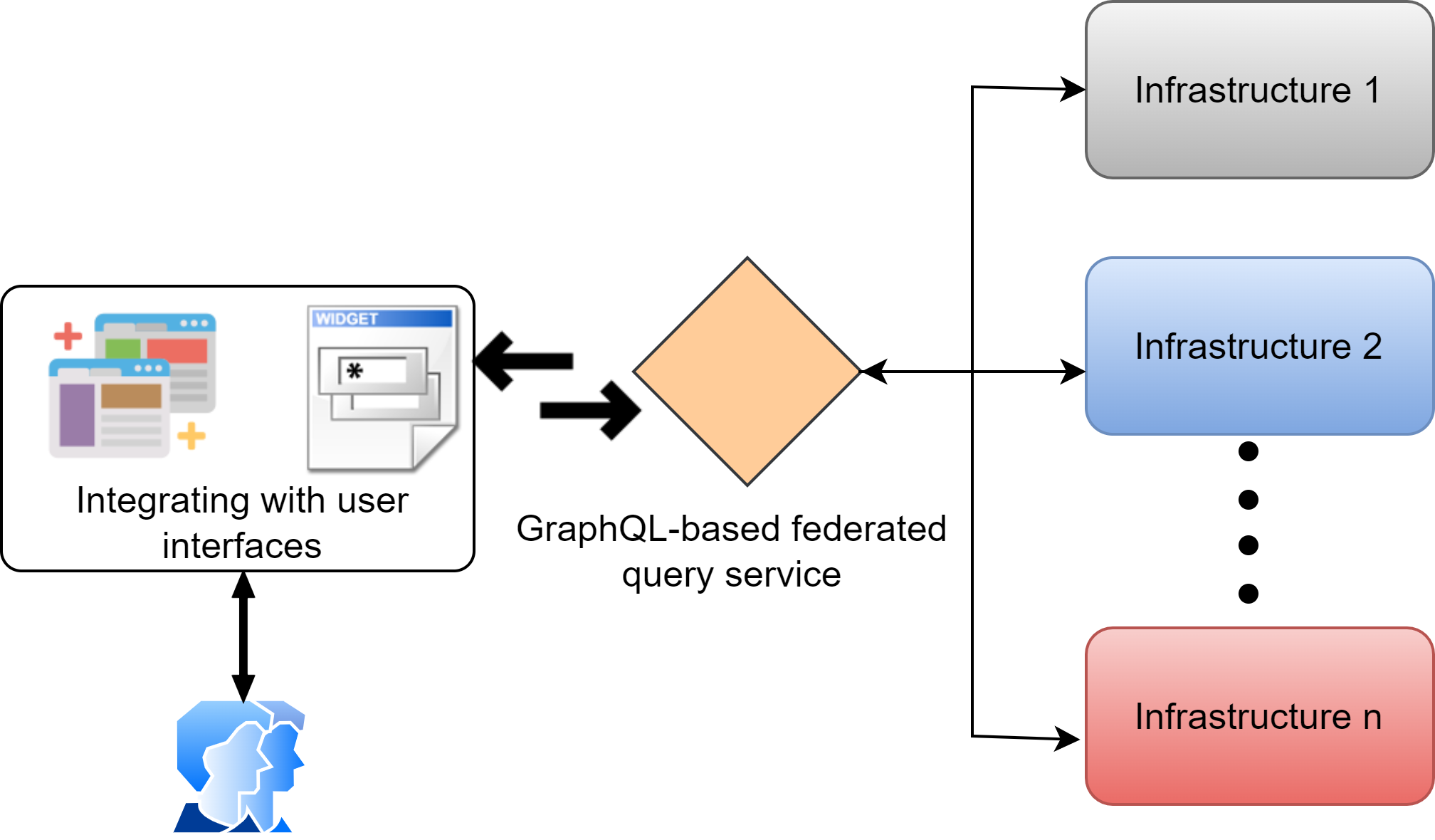}
  \caption{Conceptual model for virtually integrating numerous infrastructures to facilitate the construction of user interface widgets that enrich displayed information with context.}
  \label{fig1}
\end{figure*}

\section{Conceptual Model and its Application}
Figure~\ref{fig1} illustrates the conceptual model underpinning our work. In this model, a federated query service abstracts and unifies access to and retrieval of data served by arbitrary scholarly communication infrastructures. Here the purpose of the service is to facilitate the efficient construction of user interface widgets that enrich the presented information with contextual information. The conceptual model comprises the following two key aspects:
\begin{enumerate} 
    \item Flexible, on-demand, virtual and federated integration of scholarly communication infrastructures and straightforward extension of the GraphQL-based federated query service to serve contextual information required by user interface widgets.
    \item Uniform access by means of a single query and data exchange interface to comprehensive contextual information required to enrich with context arbitrary information presented in a user interface.
\end{enumerate}

We apply this conceptual model for scholarly communication infrastructures, specifically in developing widgets that enrich scholarly knowledge presented in the ORKG with comprehensive contextual information sourced in a federated manner from numerous scholarly communication infrastructures serving metadata about articles (Crossref and Semantic Scholar), datasets and software (DataCite), projects (OpenAIRE), organizations (ROR), contributors (ORCID). Specifically, we develop widgets to enrich scholarly knowledge presented in ORKG with rich contextual information, in particular for:
\begin{enumerate}
        \item \textit{ORKG paper view}: Display contextual information about related datasets, projects, topics and Altmetrics for the viewed paper.
        \item \textit{ORKG contributor profile}: Display employment history, published artefacts other than those published on ORKG including articles, datasets, software, projects in which the contributor was involved, and research topics of interest to the contributor.
        \item \textit{ORKG comparisons}: Extend the faceted search in ORKG comparisons with the possibility to filter the compared studies based on rich contextual metadata, e.g., filter compared studies to include those which are cited more than a given threshold.
\end{enumerate}

\section{The Federated Infrastructures} 
\label{s:background}
This section provides a brief introduction to the scholarly communication infrastructures currently included for federated data access and presents the federated query service.

\textbf{DataCite} is a DOI registration service for the persistent identification of scholarly artefacts, in particular datasets and software with a common metadata schema. The published content can be discovered in global scholarly infrastructures. DataCite also provides the PID Graph~\cite{COUSIJN2021,Fenner_Aryani_2019}, which implements the federated retrieval of metadata about and the relationships among numerous scholarly artefacts, specifically articles, datasets, software, and other entities, including organizations, projects and funders at global large-scale served by a host of scholarly communication infrastructures. The PID Graph is accessible via the DataCite GraphQL API\footnote{\url{https://api.datacite.org/graphql}}. DataCite Commons\footnote{\url{https://commons.datacite.org/}} is a web based user interface for content served by the PID Graph.

\textbf{OpenAIRE}~\cite{openairemanghi,manghi2012} enables finding and accessing scholarly articles, datasets, software, researcher profiles and information about related organization. OpenAIRE harvests metadata from multiple data providers, curates and deduplicates the metadata to provide an integrated community service. \textbf{Semantic Scholar}\footnote{\url{https://www.semanticscholar.org/}} is an AI-based web tool for searching scientific literature. Its rich REST API allows DOI-based and keyword-based queries for searching scholarly articles. \textbf{Wikidata} is a knowledge graph hosted by the Wikimedia Foundation that enables searching research articles and information about related entities (e.g. organization, people, etc.). Data available in Wikidata is accessible via REST API and SPARQL endpoint. \textbf{Altmetric}\footnote{\url{https://www.altmetric.com/}} tracks mentions of scholarly artefacts across multiple platforms, including social media. It provides a visually informative and aggregated overview of the attention research work receives online. Altmetric provides access to its data via REST API.

We integrate these scholarly communication infrastructures in a federated query service that virtually connects them at a single endpoint and enables the efficient retrieval of scholarly information in an integrated manner. The main purpose of this federation is to abstract from their heterogeneous APIs and enable virtualized, integrated access to the published content through a common unified GraphQL-based interface.

\begin{figure*}[t!]
  \includegraphics[width=\textwidth]{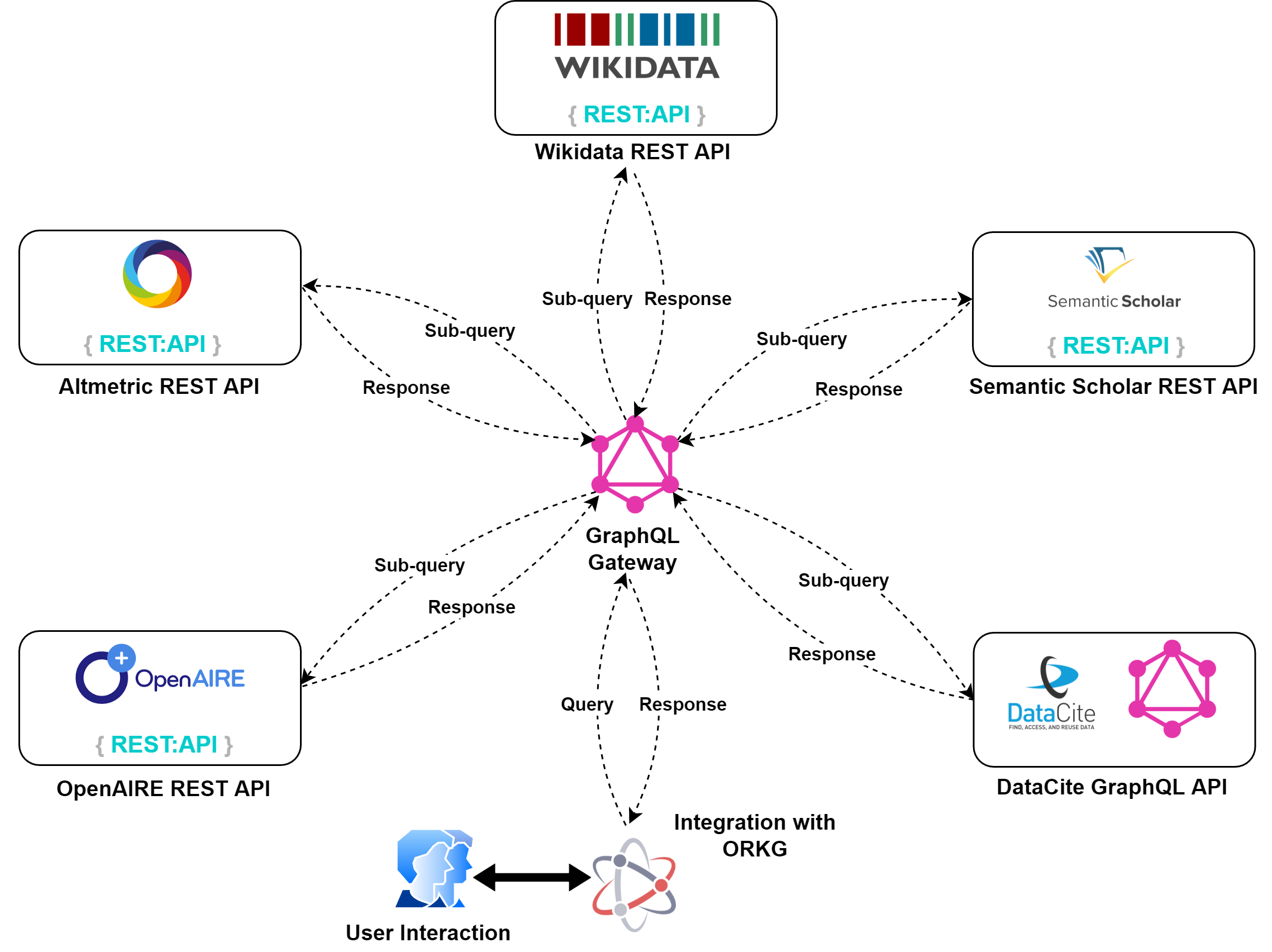}
  \caption{Overview of the virtually integrated APIs of several scholarly communication infrastructures (DataCite, OpenAIRE, Semantic Scholar, Wikidata, and Altmetric) at a GraphQL gateway, illustrating the execution of sub-queries in the respective infrastructures, and integration of the federated query service in ORKG via web widgets to retrieve and display the contextual information. Finally, the rich scholarly information is presented to the user in an aggregated form.}
  \label{fig2}
\end{figure*}

Figure~\ref{fig2} illustrates the architecture of the federated query service.This service does not contain the data itself, but implements an integrated schema for the various sources and enables the execution of queries in a federated manner. We leverage persistent identifiers for linking data served by the various infrastructures.

\begin{figure*}[t!]
  \includegraphics[width=\textwidth]{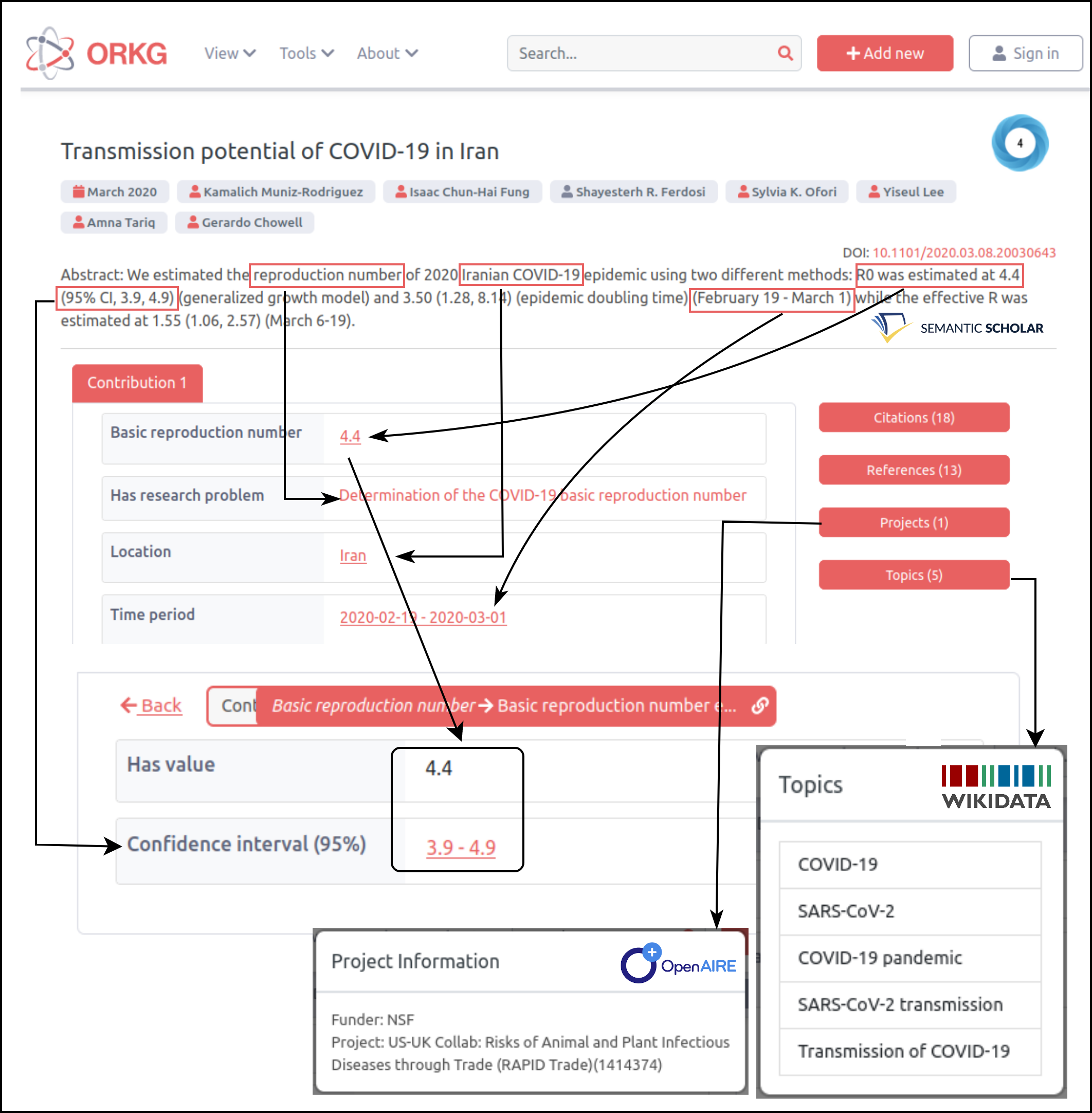}
  \caption{ORKG paper view: Fetching abstract, citations and references from Semantic Scholar; metrics data from Altmetric; project information from OpenAIRE, and related topics from Wikidata. The view also highlights how information in the article abstract is represented in a structured manner in ORKG.}
  \label{fig3}
\end{figure*}

\begin{lstfloat}
\begin{lstlisting}[language=XML, caption=Searching contextual information for the article with DOI \texttt{10.1101/2020.03.08.20030643}; Semantic Scholar provides basic metadata; OpenAIRE provides the project details; related topics are fetched from Wikidata and metrics are retrieved by querying Altmetric., basicstyle=\ttfamily\scriptsize, numbers=left, numbersep=5pt, showspaces=false, xleftmargin=0.5cm, showstringspaces=false, label=listing1, frame=tb]
{ # Semantic Scholar query
  paper(doi: "10.1101/2020.03.08.20030643") {
    doi title abstract
    citations { title doi }
    references { title doi }
    
    #OpenAIRE query
    project {
      funder project
    }
    
    #Wikidata query
    topicDetails { topic }
    
    #Altmetric query
    metricsInformation {
      url
      image
} } }
\end{lstlisting}
\end{lstfloat}

\section{Web Widgets to Enrich Knowledge with Context}
This section presents the integration of web widgets in ORKG to enrich its curated scholarly knowledge with contextual information sourced from the various scholarly communication infrastructures (see Section~\ref{s:background}). We showcase the web widget functionality for the ORKG paper view, contributor profiles, and comparison faceted search. 

\subsection{Visualizing ORKG Scholarly Knowledge with Context}
In its paper view, the ORKG presents the content of articles, i.e. the essential information contained in articles, in a structured, machine-readable form. We enrich the ORKG paper view by displaying contextual information about related datasets, projects, topics and Altmetric for the displayed article, retrieved via the described federated query service.
Figure~\ref{fig3} illustrates the ORKG paper view for an article. Upon viewing an article, the federated query service is automatically invoked through the integrated widget and requests the contextual information with a single query (see Listing~\ref{listing1}) in a federated manner. The article's meta(data) (abstract, citations, and references) is retrieved from Semantic Scholar; related projects are retrieved from OpenAIRE; Wikidata provides information about related topics; and Altmetric provides the related metrics data. The figure also highlights that the essential content published in an article is available as ORKG research contributions in structured and machine-readable form. Specifically, we highlight how some of the information contained in the article abstract obtained from Semantic Scholar (for instance, basic reproduction number and confidence interval) is represented in structured form in the ORKG. By enriching the ORKG scholarly knowledge with comprehensive contextual information we ensure that users are presented rich information thus avoiding having to explore each infrastructure individually.

\begin{figure*}[ht]
  \includegraphics[width=\textwidth]{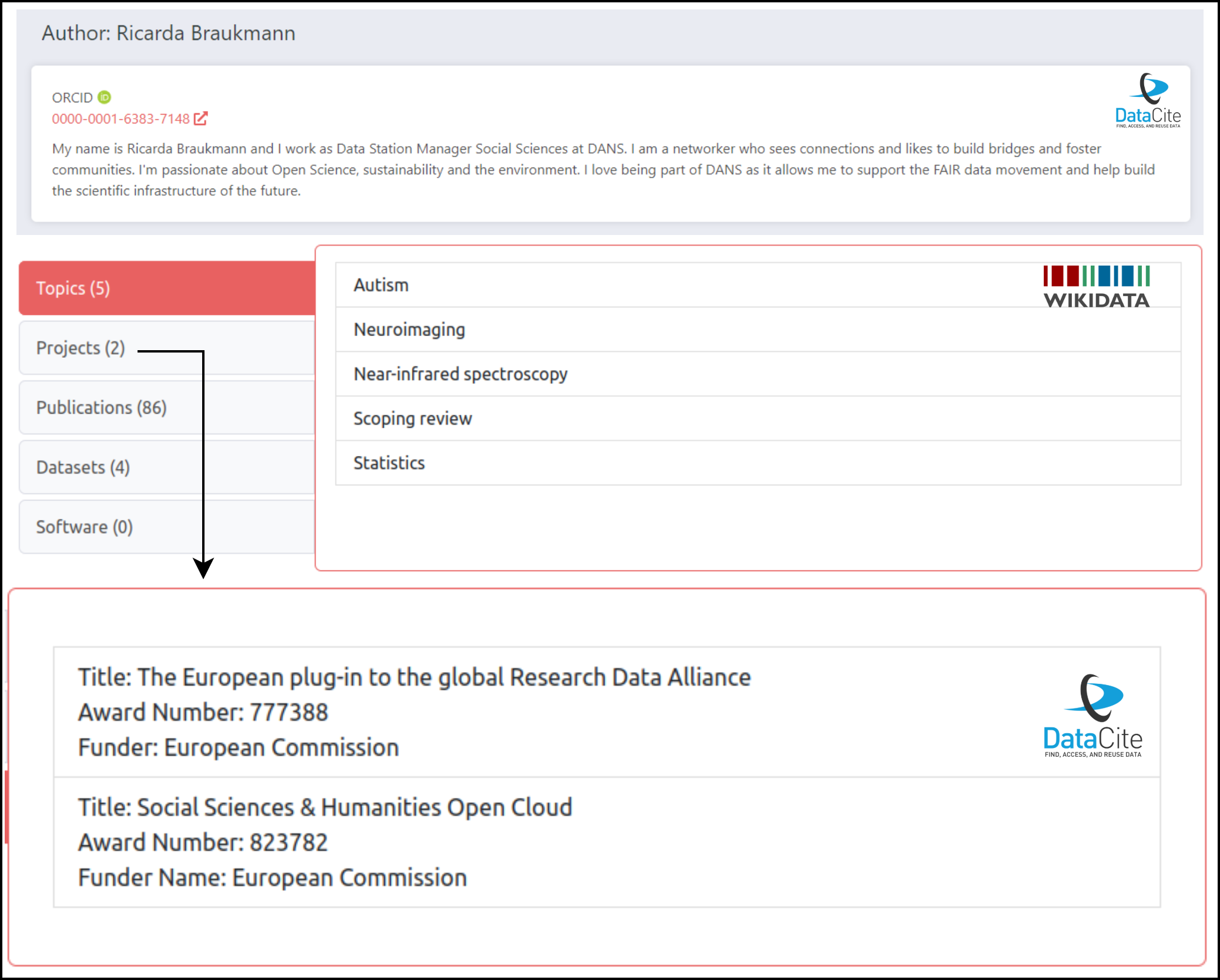}
  \caption{ORKG contributor profile for a researcher (Ricarda Braukmann) displaying employment history, projects information, research topics and other contextual information.}
  \label{fig4}
\end{figure*}

\begin{lstfloat}
\begin{lstlisting}[language=XML, caption=Federated query for retrieving a person's employment history; published scholarly artefacts; projects in which the contributor was involved in; and topics of interest., basicstyle=\ttfamily\scriptsize, numbers=left, numbersep=5pt, showspaces=false, xleftmargin=0.5cm, showstringspaces=false, label=listing2, frame=tb]
{ person(id: "https://orcid.org/0000-0001-6383-7148") {
    id name
    employment {
      organizationName
      organizationId
      startDate endDate 
    }
    publications {
      nodes { id type
        titles { title }
        fundingReferences { awardTitle awardNumber }
        creators { givenName familyName id } 
    } }
    datasets {
      totalCount
      nodes { id type
        titles { title }
        creators { givenName familyName id } 
    } }
    softwares {
      totalCount
      nodes { id type
        titles { title }
        creators { givenName familyName id } 
    } }
    topics 
} }
\end{lstlisting}
\end{lstfloat}

\subsection{Enriching ORKG Contributor Profiles}
Contributor profiles provide an overview of their work, such as published articles, datasets, software, and research topics of interest. We enrich the profile view of ORKG contributors by displaying additional contextual information along with the contributor information already available in ORKG, specifically: career history, published artefacts including articles, datasets, software as well as project involvement and research topics of interest.

Figure~\ref{fig4} shows the contextual information retrieved by ORCID of an ORKG contributor. The interface displays the employment history, published research articles, datasets, and software as well as the projects the contributor has been involved. Again, we use the federated query service to retrieve this contextual information (Listing~\ref{listing2}). The contributor's ORCID is used to retrieve publications, datasets, and software from ORCID via the PID Graph, project information from OpenAIRE, and research interests from Wikidata.

\subsection{Advanced Faceted Search for ORKG Comparisons}
ORKG comparisons are machine-readable tabular overviews of the essential content published in scholarly articles on a particular research problem~\cite{survey_literature}. These comparisons can be saved in the ORKG by specifying metadata, title, description, research field, and authors. ORKG also supports DOI-based persistent identification of comparisons to ensure their discoverability in global scholarly communication infrastructures and enable their citability. 

\begin{figure*}[ht]
  \includegraphics[width=\textwidth]{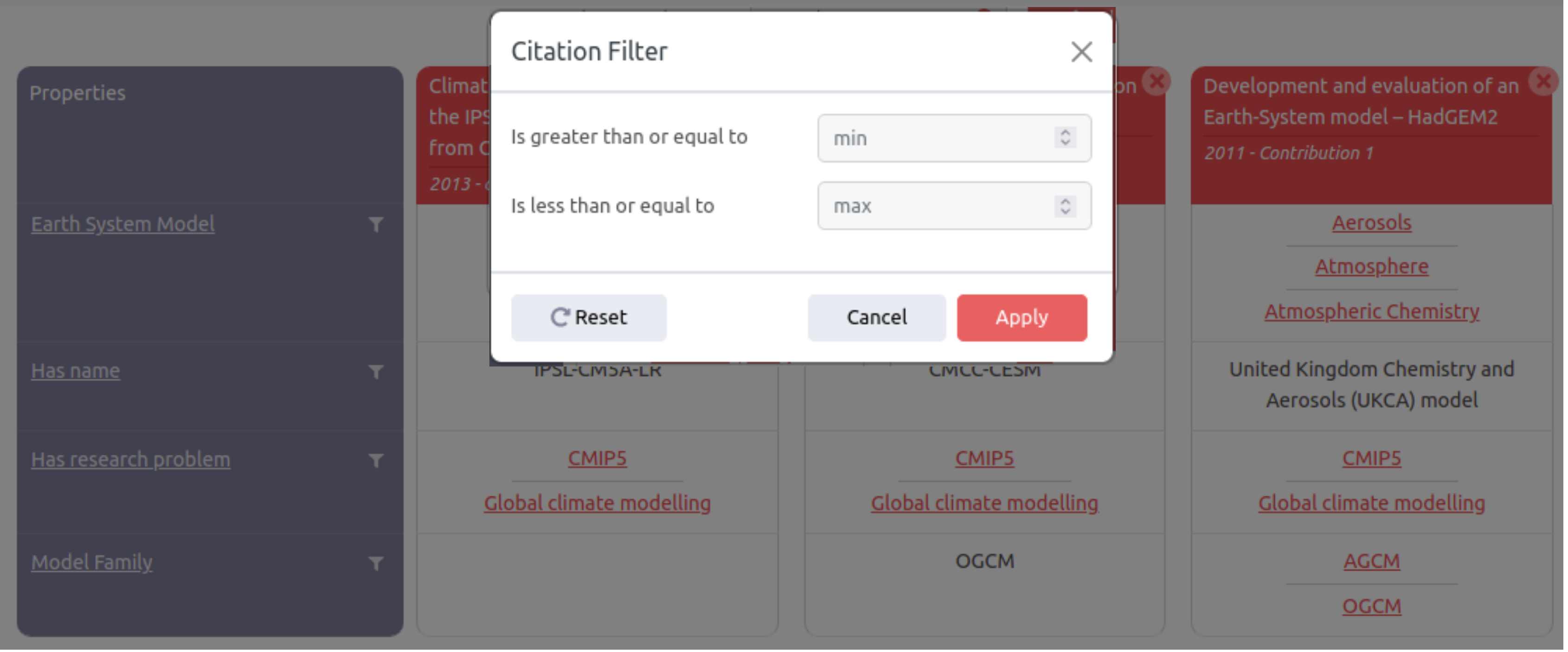}
  \caption{ORKG comparison of earth system models showing the faceted search enabling refining the compared articles by number of citations.}
  \label{fig7}
\end{figure*}

We integrate the federated query service in the ORKG comparison view to advance its faceted search functionality. Figure~\ref{fig7} shows a comparison of earth system models and a faceted search on citations to filter the comparison by articles with citations smaller or larger than given thresholds. We retrieve the number of citations for all articles included in the comparison from Semantic Scholar and refine the comparison according to the specified conditions.

\section{Discussion}
To answer our research question, we virtually integrated the DataCite PID Graph and the REST APIs of OpenAIRE, Semantic Scholar, Wikidata, and Altmetric to retrieve rich contextual information for ORKG scholarly knowledge in a federated manner, thus enabling the execution of complex distributed queries via a single gateway. The resulting data source abstraction facilitates the efficient development of web widgets that retrieve and display rich contextual information in different ORKG views for papers, contributors and comparisons. 

ORKG already supported faceted search for comparisons at content-level~\cite{Heidari_Ramadan_Stocker_Auer_2021}. The work proposed here extends this functionality with facets for contextual information and thus enables more complex (meta)data-driven filtering. For example, it is now possible to not only filter articles reporting a (COVID-19) basic reproductive number $(R0) > X$ but also having a minimum number of citations $N$. Hence, the faceted search supports filtering for specific research results that also have high impact. 
The integration of the proposed widgets in ORKG supports users in obtaining an integrated overview of scholarly knowledge and rich contextual information in a single view. 

We compared the user interfaces of ORKG, DataCite, OpenAIRE, Semantic Scholar, and Wikidata for information richness. Table~\ref{table1} shows article contextual information presented by each infrastructure. We observe that DataCite Commons and OpenAIRE present related datasets, software, and projects whereas Semantic Scholar provides information about citations and references. In contrast, ORKG presents comprehensive contextual information from these distributed scholarly infrastructures. Moreover, ORKG enables faceted search at the level of both data (i.e. article contents) and metadata (including contextual information). Lacking structured data, the other scholarly communication infrastructures are unable to provide such functionality. 

Table~\ref{table2} provides an overview of contributor contextual information presented by each infrastructure. DataCite Commons and OpenAIRE present published articles, datasets, and software, while Semantic Scholar only provides information about published articles. Wikidata also provides information about articles, including topics of interest. Compared to these infrastructures, ORKG presents a more comprehensive overview of contributor contextual information.

\begin{table*}[]
\caption{Overview of article contextual information presented by each scholarly communication infrastructure.}
\centering
\begin{tabular}{|l|l|l|l|l|l|l|l|l|}
\hline
  & \textbf{DataCite} & \textbf{OpenAIRE} & \textbf{Sem. Scholar} & \textbf{Wikidata} & \textbf{ORKG} \\ \hline
\textit{Datasets} & \cmark & \cmark & \xmark & \xmark & \cmark \\ \hline
\textit{Software} & \cmark & \cmark & \xmark & \xmark & \cmark \\ \hline
\textit{Topics} & \xmark & \xmark & \xmark & \cmark & \cmark \\ \hline
\textit{Project} & \xmark & \cmark & \xmark & \xmark & \cmark \\ \hline
\textit{Altmetric} & \xmark & \cmark & \xmark & \cmark  & \cmark \\ \hline
\textit{Citations} & partial & partial & complete & partial & complete \\ \hline
\textit{References} & partial & partial & complete & partial & complete \\ \hline
\textit{Facets} & metadata & metadata & metadata & metadata & meta/data \\ \hline
\end{tabular}
\label{table1}
\end{table*}

\begin{table*}[]
\caption{Overview of contributor contextual information presented by each scholarly communication infrastructure.}
\centering
\begin{tabular}{|l|l|l|l|l|l|l|l|}
\hline
 & \textbf{DataCite} & \textbf{OpenAIRE} & \textbf{Sem. Scholar} & \textbf{Wikidata} & \textbf{ORKG} \\ \hline
\textit{Articles} & \cmark & \cmark & \cmark & \cmark & \cmark \\ \hline
\textit{Datasets} & \cmark & \cmark & \xmark & \xmark & \cmark \\ \hline
\textit{Software} & \cmark & \cmark & \xmark & \xmark & \cmark \\ \hline
\textit{Topics} & \xmark & \xmark & \xmark & \cmark & \cmark \\ \hline
\textit{Project} & partial & \cmark & \xmark & \xmark & \cmark \\ \hline
\textit{Reviews} & \xmark & \xmark & \xmark & \xmark & \cmark \\ \hline
\end{tabular}
\label{table2}
\end{table*}

Currently, our widgets implementation focuses on articles, contributor profiles, and comparison faceted search. As the federated query service can also retrieve contextual information about organizations, we can furthermore enrich the ORKG organization view with linked projects, papers and other contextual information. This will assist users in exploring what is known about organizations, their activities, outputs, and impact.

As a further direction for future work, we will consider advancing the ORKG search with facets at both data and metadata granularity. In addition to facets for article contents, the federated query service can power facets on contextual information (primarily metadata about contextual entities). This enables users to formulate more complex requests with constraints on data and metadata. A concrete example is a search for the 10 most cited articles addressing the research problem of estimating the COVID-19 basic reproduction number that have reported a confidence interval for the estimated number less than some threshold T, and retrieve their citation count and the reported estimate for basic reproduction number of the virus.

\section{Conclusions}
We have proposed a web widget based approach for dynamic retrieval and display of comprehensive contextual information for scholarly knowledge. The approach enables rich information presentation and is powered by a GraphQL-based federated query service that virtually integrates and abstracts the technological heterogeneity of numerous scholarly communication infrastructures, in particular DataCite, OpenAIRE, Semantic Scholar, Wikidata, and Altmetric. The approach can be extended to other scholarly communication infrastructures and data sources more generally. To the best of our knowledge, no scholarly knowledge graph shows such diverse information.

As the amount of content published by scholarly communication infrastructures continues to accelerate, rich contextual information can increase research efficiency. The approach proposed and implemented in the work presented here is an important contribution towards this aim that underscores feasibility, broad applicability, and potential impact.

\section*{Acknowledgment}
This work was co-funded by the European Research Council for the project ScienceGRAPH (Grant agreement ID: 819536) and TIB--Leibniz Information Centre for Science and Technology.

\bibliographystyle{splncs04}
\bibliography{paper}

\begin{thebibliography}{10}
\providecommand{\url}[1]{\texttt{#1}}
\providecommand{\urlprefix}{URL }
\providecommand{\doi}[1]{https://doi.org/#1}

\bibitem{arya}
Arya, D., Ha-Thuc, V., Sinha, S.: Personalized federated search at linkedin.
  In: Proceedings of the 24th ACM International on Conference on Information
  and Knowledge Management. p. 1699–1702. CIKM '15, Association for Computing
  Machinery, New York, NY, USA (2015). \doi{10.1145/2806416.2806615},
  \url{https://doi.org/10.1145/2806416.2806615}

\bibitem{Scholix}
Burton, A., Koers, H., Manghi, P., Stocker, M., Fenner, M., Aryani, A.,
  La~Bruzzo, S., Diepenbroek, M., Schindler, U.: The scholix framework for
  interoperability in data-literature information exchange. D-Lib Magazine
  \textbf{23} (01 2017). \doi{10.1045/january2017-burton}

\bibitem{COUSIJN2021}
Cousijn, H., Braukmann, R., Fenner, M., Ferguson, C., {van Horik}, R., Lammey,
  R., Meadows, A., Lambert, S.: Connected research: The potential of the pid
  graph. Patterns  \textbf{2}(1),  100180 (2021).
  \doi{https://doi.org/10.1016/j.patter.2020.100180},
  \url{https://www.sciencedirect.com/science/article/pii/S2666389920302440}

\bibitem{Fenner_Aryani_2019}
Fenner, M., Aryani, A.: {Introducing the PID Graph}  (2019).
  \doi{10.5438/JWVF-8A66},
  \url{https://blog.datacite.org/introducing-the-pid-graph/}

\bibitem{federated_access}
Haris, M., Farfar, K.E., Stocker, M., Auer, S.: Federating scholarly
  infrastructures with graphql. In: Ke, H.R., Lee, C.S., Sugiyama, K. (eds.)
  Towards Open and Trustworthy Digital Societies. pp. 308--324. Springer
  International Publishing, Cham (2021)

\bibitem{haris}
Haris, M., Stocker, M.: Comparison of different scholarly communication
  infrastructures (2022). \doi{10.48366/R165794},
  \url{https://www.orkg.org/orkg/comparison/R165794}

\bibitem{Hasnain2017}
Hasnain, A., Mehmood, Q., Sana E~Zainab, S., Saleem, M., Warren, Jr, C., Zehra,
  D., Decker, S., Rebholz-Schuhman, D.: Biofed: Federated query processing over
  life sciences linked open data. Journal of Biomedical Semantics  \textbf{8}
  (03 2017). \doi{10.1186/s13326-017-0118-0}

\bibitem{Heibi2019}
Heibi, I., Peroni, S., Shotton, D.: Enabling text search on sparql endpoints
  through oscar. Data Science  \textbf{2} (04 2019). \doi{10.3233/DS-190016}

\bibitem{Heidari_Ramadan_Stocker_Auer_2021}
Heidari, G., Ramadan, A., Stocker, M., Auer, S.: Leveraging a federation of
  knowledge graphs to improve faceted search in digital libraries (2021).
  \doi{10.1007/978-3-030-86324-1\_18}

\bibitem{himmelstein2017systematic}
Himmelstein, D.S., Lizee, A., Hessler, C., Brueggeman, L., Chen, S.L., Hadley,
  D., Green, A., Khankhanian, P., Baranzini, S.E.: Systematic integration of
  biomedical knowledge prioritizes drugs for repurposing. Elife  \textbf{6},
  e26726 (2017)

\bibitem{orkg}
Jaradeh, M.Y., Oelen, A., Farfar, K.E., Prinz, M., D'Souza, J., Kismih\'{o}k,
  G., Stocker, M., Auer, S.: Open research knowledge graph: Next generation
  infrastructure for semantic scholarly knowledge. In: 10th Int. Conf. on
  Knowledge Capture. K-CAP '19, ACM (2019). \doi{10.1145/3360901.3364435}

\bibitem{khan2017survey}
Khan, S., Liu, X., Shakil, K.A., Alam, M.: A survey on scholarly data: From big
  data perspective. Information Processing \& Management  \textbf{53}(4),
  923--944 (2017)

\bibitem{Kurteva2021}
Kurteva, A., De~Ribaupierre, H.: Interface to query and visualise definitions
  from a knowledge base. In: Brambilla, M., Chbeir, R., Frasincar, F.,
  Manolescu, I. (eds.) Web Engineering. pp. 3--10. Springer (2021)

\bibitem{liekens2011biograph}
Liekens, A.M., De~Knijf, J., Daelemans, W., Goethals, B., De~Rijk, P.,
  Del-Favero, J.: Biograph: unsupervised biomedical knowledge discovery via
  automated hypothesis generation. Genome biology  \textbf{12}(6),  1--12
  (2011)

\bibitem{openairemanghi}
Manghi, P., Bolikowski, L., Manola, N., Schirrwagen, J., Smith, T.:
  Openaireplus: the european scholarly communication data infrastructure. D-Lib
  Magazine  \textbf{18} (09 2012). \doi{10.1045/september2012-manghi}

\bibitem{manghi2012}
Manghi, P., Houssos, N., Mikulicic, M., J{\"o}rg, B.: The data model of the
  openaire scientific communication e-infrastructure. In: Dodero, J.M.,
  Palomo-Duarte, M., Karampiperis, P. (eds.) Metadata and Semantics Research.
  pp. 168--180. Springer Berlin Heidelberg, Berlin, Heidelberg (2012)

\bibitem{Morton2019}
Morton, K., Wang, P., Bizon, C., Cox, S., Balhoff, J., Kebede, Y., Fecho, K.,
  Tropsha, A.: {ROBOKOP: an abstraction layer and user interface for knowledge
  graphs to support question answering}. Bioinformatics  \textbf{35}(24),
  5382--5384 (08 2019). \doi{10.1093/bioinformatics/btz604},
  \url{https://doi.org/10.1093/bioinformatics/btz604}

\bibitem{federatedengine}
Mosharraf, M., Taghiyareh, F.: Federated search engine for open educational
  linked data. Bull. IEEE Tech. Comm. Learn. Technol  \textbf{18}(6) (2016)

\bibitem{Nielsen2017}
Nielsen, F.{\AA}., Mietchen, D., Willighagen, E.: Scholia, scientometrics and
  wikidata. In: Blomqvist, E., Hose, K., Paulheim, H., {\L}awrynowicz, A.,
  Ciravegna, F., Hartig, O. (eds.) The Semantic Web: ESWC 2017 Satellite
  Events. pp. 237--259. Springer International Publishing, Cham (2017)

\bibitem{survey_literature}
Oelen, A., Jaradeh, M.Y., Stocker, M., Auer, S.: Generate fair literature
  surveys with scholarly knowledge graphs. In: Proceedings of the ACM/IEEE
  Joint Conference on Digital Libraries in 2020. p. 97–106. JCDL '20,
  Association for Computing Machinery, New York, NY, USA (2020).
  \doi{10.1145/3383583.3398520}

\bibitem{safder2018ai}
Safder, I., Hassan, S.U., Aljohani, N.R.: Ai cognition in searching for
  relevant knowledge from scholarly big data, using a multi-layer perceptron
  and recurrent convolutional neural network model. In: Companion Proceedings
  of the The Web Conference 2018. pp. 251--258 (2018)

\bibitem{Schwarte2011FedXOT}
Schwarte, A., Haase, P., Hose, K., Schenkel, R., Schmidt, M.: Fedx:
  Optimization techniques for federated query processing on linked data. In:
  International Semantic Web Conference (2011)

\bibitem{Stocker_instruments}
Stocker, M., Darroch, L., Krahl, R., Habermann, T., Devaraju, A., Schwardmann,
  U., D'Onofrio, C., H{\"a}ggstr{\"o}m, I.: Persistent identification of
  instruments. Data Science Journal  \textbf{19},  1--12 (05 2020).
  \doi{10.5334/dsj-2020-018}

\bibitem{xia2017big}
Xia, F., Wang, W., Bekele, T.M., Liu, H.: Big scholarly data: A survey. IEEE
  Transactions on Big Data  \textbf{3}(1),  18--35 (2017)

\bibitem{zaki2017}
Zaki, N., Tennakoon, C.: Biocarian: Search engine for exploratory searches in
  heterogeneous biological databases. BMC Bioinformatics  \textbf{18}, ~435 (10
  2017). \doi{10.1186/s12859-017-1840-4}

\bibitem{6666765}
{Zhou}, Y., {De}, S., {Moessner}, K.: Implementation of federated query
  processing on linked data. In: 2013 IEEE 24th Annual International Symposium
  on Personal, Indoor, and Mobile Radio Communications (PIMRC). pp. 3553--3557
  (2013). \doi{10.1109/PIMRC.2013.6666765}

\end{thebibliography}
\end{document}